\documentclass[useAMS,usenatbib,a4paper,referee]{mn2e}
\usepackage{graphicx}
\usepackage{amsmath}
\usepackage{txfonts}
\usepackage{mathrsfs}
\usepackage{longtable}

\def\aap{A\&A}
\def\apjl{ApJ}

\def\apjs{ApJS}

\newcommand{\be}{\begin{equation}}
\newcommand{\ee}{\end{equation}}
\newcommand{\bea}{\begin{eqnarray}}
\newcommand{\eea}{\end{eqnarray}}

\title[Strong-lensing Systems]{Model Selection with Strong-lensing Systems}
\author[Leaf \& Melia]{Kyle Leaf$^{1}$\thanks{kyleaf@email.arizona.edu} and
Fulvio Melia$^{2}$\thanks{John Woodruff Simpson Fellow. E-mail: fmelia@email.arizona.edu} \\
$^1$Department of Physics, The University of Arizona, AZ 85721, USA \\
$^2$Department of Physics, The Applied Math Program, and Department of Astronomy, 
The University of Arizona, AZ 85721, USA}

\begin{document}

\date{}

\pagerange{\pageref{firstpage}--\pageref{lastpage}} \pubyear{2016}

\maketitle

\label{firstpage}

\begin{abstract}
In this paper, we use an unprecedentedly large sample (158)
of confirmed strong lens systems for model selection, comparing
five well studied Friedmann-Robertson-Walker cosmologies:
$\Lambda$CDM, $w$CDM (the standard model with a variable dark-energy equation
of state), the $R_{\rm h}=ct$ universe, the (empty) Milne cosmology,
and the classical Einstein-de Sitter (matter dominated) universe. 
We first use these sources to optimize the parameters in the standard
model and show that they are consistent with {\it Planck}, though
the quality of the best fit is not satisfactory. We demonstrate that
this is likely due to under-reported errors, or to errors yet to be
included in this kind of analysis. We suggest that the missing
dispersion may be due to scatter about a pure single isothermal
sphere (SIS) model that is often assumed for the mass distribution in
these lenses. We then use the Bayes information criterion, with the
inclusion of a suggested SIS dispersion, to calculate the relative 
likelihoods and ranking of these models, showing that Milne and 
Einstein-de Sitter are completely ruled out, while $R_{\rm h}=ct$ is 
preferred over $\Lambda$CDM/$w$CDM with a relative probability of 
$\sim 73\%$ versus $\sim 24\%$. The recently reported sample of 
new strong lens candidates by the Dark Energy Survey, if confirmed, may 
be able to demonstrate which of these two models is favoured over the 
other at a level exceeding $3\sigma$.
\end{abstract}


\newpage
\section{Introduction}
With the continued discovery of new strong lensing systems, the gravitational 
bending of light is gaining in importance as a diagnostic tool for the expansion
of the Universe. Einstein's initial conclusion regarding the subject of gravitational 
lensing was that such a phenomenon would be very difficult to observe (Einstein 1936).
It would take another sixty years before the first complete Einstein Ring was discovered 
(King et al. 1998). Strong gravitational lenses exist as uniquely geometrical phenomena, 
dependent only on the mass distribution of the nearer object and the distances between 
observer, lens, and source. Therefore, they offer a probe of the expansion history from 
the time the source emitted its light to the present day. 

There has been a rapid progression in the number of known strongly lensed systems
since the turn of the century, starting with The Lenses Structure and Dynamics (LSD) 
survey (Koopmans \& Treu 2003; Treu \& Koopmans 2002, 2004). Its successor, The Sloan 
Lens ACS (Advanced Camera for Surveys) (SLACS) found an additional 57 confirmed lenses 
(Bolton et al. 2008; Auger et al. 2009). Most recently, the SLACS survey has discovered 
an additional 40 lenses (Shu et al. 2017). The Baryon Oscillation Spectroscopic Survey 
(BOSS; Brownstein et al. 2012) produced another 25 confirmed strong galaxy-galaxy lenses, 
and the Strong Lensing Legacy Survey (SL2S; Gavazzi et al. 2014; Sonnenfeld et al. 2013a,2013b),
has resulted in a catalog of 31 confirmed lenses. Most recently, the Dark Energy Survey has 
compiled a catalog of 374 candidate systems (several of which were identified in prior 
surveys) awaiting follow-up observations for confirmation in the near future (Diehl et al. 
2017). Combined, these surveys include lenses from redshifts $z=0.06$ to $1$, and sources
from $z=0.2$ to $3.6$ ($9$ of them beyond $z=3$). This rapid expansion in the catalog
of available lens systems is crucial for cosmological work because the impact from
the testing of cosmological models using these objects depends heavily on both the
number of observations and the redshift range of the set. 

In this paper, we use a catalog of 158 confirmed, strong lens systems suitable for 
testing various expansion scenarios---a significantly larger compilation than that
of our previous analysis (Melia, Wei \& Wu 2015) and that of Cao et al. (2015). 
We begin by constraining the properties of dark energy within $w$CDM (i.e., the
standard model with a variable dark-energy equation of state), and then proceed
to use model selection tools to determine which of several models is preferred 
by the strong-lens data. We consider the $R_{\rm h}=ct$ universe, a Friedmann-Robertson-Walker
(FRW) cosmology with zero active mass, i.e., $\rho+3p=0$, in terms of the total energy
density $\rho$ and pressure $p$ (Melia 2007, 2016, 2017a; Melia \& Abdelqader 2009;
Melia \& Shevchuk 2012), the matter only Einstein-de-Sitter universe, and the 
empty Milne model (see, e.g., Vishwakarma 2013). 

Our previous analysis (Melia, Wei, \& Wu 2015) used a catalog of 69 sources and found 
that $\Lambda$CDM and $R_h=ct$ both performed reasonably well, though the number of
measurements was insufficient to favor one model over the other. We also constructed
a much larger mock catalog to estimate how many lensing systems would be required to 
carry out a definitive model selection, and concluded that a sample of several hundred 
lenses would suffice. The catalog of 158 systems we use here is approaching this 
threshold. But note that several of the 69 sources used in our previous study
are not included here. In this paper, we restrict our attention to galaxy-galaxy 
lens systems, while some in the (Melia, Wei, \& Wu 2015) analysis are of other types.
When performing a comparison between the $R_{\rm h}=ct$ universe and 
$\Lambda$CDM, however, it is important to recognize that these models make very similar 
predictions at low redshifts, meaning that the most important sources in this model 
selection are those at the highest redshifts. Unfortunately, the most recent additions 
from the SLACS survey (Shu et al. 2017) constitute only sources with a measured redshift 
$z\lesssim 1.3$. While these lenses are indeed useful for verifying the low-redshift 
predictions of any model, and for constraining the parameters of $w$CDM, they do not 
significantly contribute to the model selection itself. The SL2S and LSD surveys 
include all considered sources of redshift $z\gtrsim 1.6$, and therefore have the 
greatest impact on a direct comparison of the models. In \S~2, we present the methods 
used to perform the fitting and model comparison. In \S~3 we discuss the results of 
these calculations, and we end with a summary and our conclusions in \S~4.

\section{Strong Lensing}
Strong gravitational lensing has been used to constrain cosmological models in 
several recent publications, including Cao et al. (2012, 2015), Melia, Wei \& Wu 
(2015), and An, Chang \& Xu (2016). For a strongly lensed system with a single 
galaxy acting as the lens, the Einstein Radius depends only on three parameters:
the angular diameter distances to the lens and source, and the mass distribution 
within the lensing galaxy. The most commonly used model for the lens galaxy's mass 
is a singular isothermal ellipsoid (SIE) (Ratnatunga, Griffiths \& Ostrander 1999). 
Early-type galaxies (ellipticals) contain most of the cosmic stellar mass in the 
universe, and are therefore more commonly found as lenses than other types; 
approximating these as an ellipsoid is quite reasonable (Kochanek et al. 2000). 
The prior analyses by Cao et al. (2015) and Melia, Wei, \& Wu (2015), however, 
reported consistency also with the simpler Singular Isothermal Sphere (SIS) 
model (i.e., an SIE with zero ellipticity). The transparency of the results and
the simplicity of the methodology therefore warrant making the SIS approximation.
Nonetheless, to ensure that our results are not biased by this approach, we also
consider potential random deviations from this simple model. An SIS determines 
the Einstein (angular) radius $\theta_E$ in terms of the 1-D velocity dispersion,
$\sigma_{SIS}$, in the lensing galaxy. In terms of the ratio 
\begin{equation}
\mathcal{D}=\dfrac{{D}_{ls}}{{D}_{s}}\;,
\end{equation}
of the angular diameter distances $D_{ls}$ and $D_s$ between the lens and source, 
and source and observer, respectively, the Einstein radius is given as
\begin{equation}
\theta_E=4{\pi}\dfrac{\sigma_{SIS}^2}{c^2}\,\mathcal{D}\;.
\end{equation}

Note, however, that in place of $\sigma_{SIS}$, we follow the approach taken by
Cao et al. (2015) in converting the observed velocity dispersion $\sigma_{\rm ap}$
measured within a given aperture and convert it to a velocity dispersion within a 
circular aperture of half the effective radius of the lens galaxy, with 
$\sigma_0=\sigma_{\rm ap}(\theta_{\rm eff}/2\theta_{\rm ap})^{-0.04}$, where $\theta_{\rm eff}$ 
is the half-light radius of the lensing galaxy and $\theta_{\rm ap}$ is the aperture
size used to measure the velocity dispersion (J{\o}rgensen, Franx \& Kj{\ae}rgaard 1995a, 1995b).
The ratio $\mathcal{D}$ therefore constitutes an observable quantity, written as 
\begin{equation}
\mathcal{D}^{\rm obs}=\dfrac{c^2\theta_E}{4{\pi}\,\sigma_0^2}.
\end{equation}

The quantity $\mathcal{D}$ is well-defined for any given cosmology if the redshifts 
of the lens and source are known. In flat $w$CDM, we have
\begin{equation}
D{(z_1,z_2)}=\dfrac{c}{H_0}\dfrac{1}{1+z_2}\int_{z_1}^{z_2}\dfrac{dz^\prime}
{\sqrt{\Omega_{\rm m}(1+z^\prime)^3+\Omega_{\rm r}(1+z^\prime)^4+\Omega_{\rm de}
(1+z^\prime)^{3(1+w_{\rm de})}}}\;,
\end{equation}
where $w_{\rm de}\equiv p_{\rm de}/\rho_{\rm de}$ is the dark-energy equation of
state parameter, and $\Omega_{i}$ is today's density of species $i$ in terms of
the critical density $\rho_{\rm c}\equiv 3c^2 H_0^2/8\pi G$. For our strong lens 
sample, $\Omega_{\rm r}$ is negligible and may be ignored. For flatness, we also 
have $\Omega_{\rm m}+\Omega_{\rm de}=1$, consistent with the latest {\it Planck}
data release (Planck Collaboration 2016). This leaves only three free parameters,
$H_0$, $w_{\rm de}$ and $\Omega_{\rm m}$. When taking the ratio of two angular
diameter distances, however, $H_0$ itself becomes irrelevant, so fitting the 
strong lens sample with $w$CDM reduces to an optimization based on only two free
parameters. 

We may use the same expression (Eq.~4) for the Einstein-De-Sitter universe, 
setting $\Omega_{\rm m}$=1, $\Omega_{\rm r}=0$ and $\Omega_{\rm de}=0$.
For the Milne universe, the corresponding expression is
\begin{equation}
D{(z_1,z_2)}=\dfrac{c}{H_0}\dfrac{1}{1+z_2}\sinh\left(\ln
\left[\dfrac{1+z_2}{1+z_1}\right]\right)
\end{equation}
(e.g., Vishwakarma 2015), while for the $R_{\rm h}=ct$ model it is simply
\begin{equation}
D{(z_1,z_2)}=\dfrac{c}{H_0}\dfrac{1}{1+z_2}\ln\left[\dfrac{1+z_2}{1+z_1}\right]
\end{equation}
(e.g., Melia, Wei \& Wu 2015). The model predictions compared to the data
are based on the expected ratio $\mathcal{D}^{\rm th}$, defined in general as
\begin{equation}
\mathcal{D}^{\rm th}=\dfrac{D(z_1,z_2)}{D(0,z_2)}.
\end{equation}
For example, in the $R_{\rm h}=ct$ universe, it is 
\begin{equation}
\mathcal{D}^{\rm th}{(z_1,z_2)}=1-\dfrac{\ln(1+z_1)}{\ln(1+z_2)}\;,
\end{equation}
with corresponding expressions for the other cosmologies. Clearly, 
$\mathcal{D}$ does not depend on $H_0$ for any of the considered models, 
removing all free parameters in $R_{\rm h}=ct$, Milne, and Einstein-De-Sitter, 
while leaving $w$CDM with the two free parameters, $w_{\rm de}$ and 
$\Omega_{\rm m}$, for the optimization process. Note also that, by definition,
this ratio is restricted to the range $0\le\mathcal{D}\le 1$ for all lens systems.

\section{Data and Methodology}
Previous analyses of strong lenses used a variety of approaches to constrain the 
model parameters and for model selection (Melia, Wei, \& Wu 2015; Cao et al. 2015; 
An, Chang \& Xu 2016). For example, Cao et al. (2015) used a subset of the data we 
consider in this paper and several statistical features that warrant further consideration. 
In this paper, we compile a catalog of 158 confirmed sources (see Table~1), many 
identical to those included in Cao et al. (2015), but with the addition of 40 more 
discovered by SLACS (Shu et al. 2017). All redshifts are determined spectroscopically, 
and we use the Einstein Radii measured by the discovery teams based on fits to 
pixelized images of the sources. Cao et al. (2015) found that---assuming the {\it Planck} 
optimized value for $\Omega_{\rm m}$---the $w$CDM model is consistent to within $1\sigma$
with flat $\Lambda$CDM. Their fitting utilized the errors reported by the 
various surveys, in addition to assuming a uniform error of $5\%$ for the measured 
Einstein Radius $\sigma_{\theta_E}$ (Grillo, Lombardi \& Bertin, 2008).
The expression for the combined error in $\mathcal{D}^{\rm obs}$ is then 
\begin{equation}
\sigma_\mathcal{D}=\sqrt{4(\sigma_{\sigma_0})^2+(\sigma_{\theta_E})^2},
\end{equation}
where $\sigma_{\sigma_0}$ is the error reported for the velocity dispersion. 
Note that, while $\sigma_0$ does depend on the measured effective radius 
$\theta_{\rm eff}$, this is also determined to better than $5\%$ accuracy, 
and the low power index of $0.04$ (see expression following Eq.~2 above)
results in an insignificant error contribution compared to that from the 
velocity dispersion itself. 

Some lensing systems have two images, while others have four, a distinction
that could generate some systematic differences between the two sub-groups. 
The previous analysis by Melia, Wei, \& Wu (2015), however, showed that
there are no significant differences between two-image and four-image systems.
Given (i) that the recent SLACS data are not characterized in terms of which 
sub-group they belong to, and (ii) that there does not appear to be any dependence
of the analysis on the number of images, we do not consider the two sub-samples
separately here.

\begin{center}
\begin{longtable}{lccrccccr}
\caption{Strong-lensing Systems}\\
\hline\hline
Name &  $z_l$ &  $z_s$ &  $\sigma_{ap}\quad$ &  $\theta_E$ & Survey & $\theta_{ap}$ ('') & $\theta_{eff}$ & $\sigma_0\quad$ \\
&&&(km s$^{-1}$)&(arcsec)&&(arcsec)&(arcsec)&(km s$^{-1}$)\\
\hline
&&\\
J0151+0049 & 0.517 & 1.364 & 219$\pm$39 & 0.68 & BELLS & 1.00 & 0.89 & 226$\pm$40 \\
J0747+4448 & 0.437 & 0.897 & 281$\pm$52 & 0.61 & BELLS & 1.00 & 1.24 & 286$\pm$53 \\
J0747+5055 & 0.438 & 0.898 & 328$\pm$60 & 0.75 & BELLS & 1.00 & 2.87 & 323$\pm$59 \\
J0801+4727 & 0.483 & 1.518 & 98$\pm$24 & 0.49 & BELLS & 1.00 & 0.57 & 103$\pm$25 \\
J0830+5116 & 0.530 & 1.332 & 268$\pm$36 & 1.14 & BELLS & 1.00 & 1.10 & 274$\pm$37 \\
J0944-0147 & 0.539 & 1.179 & 204$\pm$34 & 0.72 & BELLS & 1.00 & 1.35 & 207$\pm$35 \\
J1159-0007 & 0.579 & 1.346 & 165$\pm$41 & 0.68 & BELLS & 1.00 & 0.99 & 170$\pm$42 \\
J1215+0047 & 0.642 & 1.297 & 262$\pm$45 & 1.37 & BELLS & 1.00 & 1.42 & 266$\pm$46 \\
J1221+3806 & 0.535 & 1.284 & 187$\pm$48 & 0.70 & BELLS & 1.00 & 0.93 & 193$\pm$49 \\
J1234-0241 & 0.490 & 1.016 & 122$\pm$31 & 0.53 & BELLS & 1.00 & 1.61 & 123$\pm$31 \\
J1318-0104 & 0.659 & 1.396 & 177$\pm$27 & 0.68 & BELLS & 1.00 & 1.06 & 182$\pm$28 \\
J1337+3620 & 0.564 & 1.182 & 225$\pm$35 & 1.39 & BELLS & 1.00 & 1.60 & 227$\pm$35 \\
J1349+3612 & 0.440 & 0.893 & 178$\pm$18 & 0.75 & BELLS & 1.00 & 2.03 & 178$\pm$18 \\
J1352+3216 & 0.463 & 1.034 & 161$\pm$21 & 1.82 & BELLS & 1.00 & 1.35 & 164$\pm$21 \\
J1522+2910 & 0.555 & 1.311 & 166$\pm$27 & 0.74 & BELLS & 1.00 & 1.08 & 170$\pm$28 \\
J1541+1812 & 0.560 & 1.113 & 174$\pm$24 & 0.64 & BELLS & 1.00 & 0.59 & 183$\pm$25 \\
J1542+1629 & 0.352 & 1.023 & 210$\pm$16 & 1.04 & BELLS & 1.00 & 1.45 & 213$\pm$16 \\
J1545+2748 & 0.522 & 1.289 & 250$\pm$37 & 1.21 & BELLS & 1.00 & 2.65 & 247$\pm$37 \\
J1601+2138 & 0.544 & 1.446 & 207$\pm$36 & 0.86 & BELLS & 1.00 & 0.63 & 217$\pm$38 \\
J1611+1705 & 0.477 & 1.211 & 109$\pm$23 & 0.58 & BELLS & 1.00 & 1.33 & 111$\pm$23 \\
J1631+1854 & 0.408 & 1.086 & 272$\pm$14 & 1.63 & BELLS & 1.00 & 2.07 & 272$\pm$14 \\
J1637+1439 & 0.391 & 0.874 & 208$\pm$30 & 0.65 & BELLS & 1.00 & 0.89 & 215$\pm$31 \\
J2122+0409 & 0.626 & 1.452 & 324$\pm$56 & 1.58 & BELLS & 1.00 & 1.76 & 326$\pm$56 \\
J2125+0411 & 0.363 & 0.978 & 247$\pm$17 & 1.20 & BELLS & 1.00 & 1.47 & 250$\pm$17 \\
J2303+0037 & 0.458 & 0.936 & 274$\pm$31 & 1.02 & BELLS & 1.00 & 1.35 & 278$\pm$31 \\
CFRS03-1077 & 0.938 & 2.941 & 251$\pm$19 & 1.24 & LSD & 1.25 & 1.60 & 256$\pm$19 \\
HST-14176 & 0.810 & 3.399 & 224$\pm$15 & 1.41 & LSD & 1.25 & 1.06 & 232$\pm$16 \\
HST-15433 & 0.497 & 2.092 & 116$\pm$10 & 0.36 & LSD & 1.25 & 0.41 & 125$\pm$11 \\
MG-2016 & 1.004 & 3.263 & 328$\pm$32 & 1.56 & LSD & 0.65 & 0.31 & 347$\pm$34 \\
Q0047-2808 & 0.485 & 3.595 & 229$\pm$15 & 1.34 & LSD & 1.25 & 0.82 & 239$\pm$16 \\
J0212-0555 & 0.750 & 2.740 & 273$\pm$22 & 1.27 & SL2S & 0.90 & 1.22 & 277$\pm$22 \\
J0213-0743 & 0.717 & 3.480 & 293$\pm$34 & 2.39 & SL2S & 1.00 & 1.97 & 293$\pm$34 \\
J0214-0405 & 0.609 & 1.880 & 287$\pm$47 & 1.41 & SL2S & 1.00 & 1.21 & 293$\pm$48 \\
J0217-0513 & 0.646 & 1.847 & 239$\pm$27 & 1.27 & SL2S & 1.50 & 0.73 & 253$\pm$29 \\
J0219-0829 & 0.389 & 2.150 & 289$\pm$23 & 1.30 & SL2S & 1.00 & 0.95 & 298$\pm$24 \\
J0223-0534 & 0.499 & 1.440 & 288$\pm$28 & 1.22 & SL2S & 1.00 & 1.31 & 293$\pm$28 \\
J0225-0454 & 0.238 & 1.199 & 234$\pm$21 & 1.76 & SL2S & 1.00 & 2.12 & 233$\pm$21 \\
J0226-0420 & 0.494 & 1.232 & 263$\pm$24 & 1.19 & SL2S & 1.00 & 0.84 & 272$\pm$25 \\
J0232-0408 & 0.352 & 2.340 & 281$\pm$26 & 1.04 & SL2S & 1.00 & 1.14 & 287$\pm$27 \\
J0848-0351 & 0.682 & 1.550 & 197$\pm$21 & 0.85 & SL2S & 0.90 & 0.45 & 208$\pm$22 \\
J0849-0251 & 0.274 & 2.090 & 276$\pm$35 & 1.16 & SL2S & 0.90 & 1.34 & 279$\pm$35 \\
J0849-0412 & 0.722 & 1.540 & 320$\pm$24 & 1.10 & SL2S & 0.90 & 0.46 & 338$\pm$25 \\
J0850-0347 & 0.337 & 3.250 & 290$\pm$24 & 0.93 & SL2S & 0.70 & 0.28 & 309$\pm$26 \\
J0855-0147 & 0.365 & 3.390 & 222$\pm$25 & 1.03 & SL2S & 0.70 & 0.69 & 228$\pm$26 \\
J0855-0409 & 0.419 & 2.950 & 281$\pm$22 & 1.36 & SL2S & 0.70 & 1.13 & 283$\pm$22 \\
J0904-0059 & 0.611 & 2.360 & 183$\pm$21 & 1.40 & SL2S & 0.90 & 2.00 & 182$\pm$21 \\
J0959+0206 & 0.552 & 3.350 & 188$\pm$22 & 0.74 & SL2S & 0.90 & 0.46 & 199$\pm$23 \\
J1359+5535 & 0.783 & 2.770 & 228$\pm$29 & 1.14 & SL2S & 1.00 & 1.13 & 233$\pm$30 \\
J1404+5200 & 0.456 & 1.590 & 342$\pm$20 & 2.55 & SL2S & 1.00 & 2.03 & 342$\pm$20 \\
J1405+5243 & 0.526 & 3.010 & 284$\pm$21 & 1.51 & SL2S & 1.00 & 0.83 & 294$\pm$22 \\
J1406+5226 & 0.716 & 1.470 & 253$\pm$19 & 0.94 & SL2S & 1.00 & 0.80 & 262$\pm$20 \\
J1411+5651 & 0.322 & 1.420 & 214$\pm$23 & 0.93 & SL2S & 1.00 & 0.85 & 221$\pm$24 \\
J1420+5258 & 0.38 & 0.990 & 246$\pm$23 & 0.96 & SL2S & 1.00 & 1.11 & 252$\pm$24 \\
J1420+5630 & 0.483 & 3.120 & 228$\pm$19 & 1.40 & SL2S & 1.00 & 1.62 & 230$\pm$19 \\
J2203+0205 & 0.400 & 2.150 & 213$\pm$21 & 1.95 & SL2S & 1.00 & 0.99 & 219$\pm$22 \\
J2205+0147 & 0.476 & 2.530 & 317$\pm$30 & 1.66 & SL2S & 0.90 & 0.66 & 330$\pm$31 \\
J2213-0009 & 0.338 & 3.450 & 165$\pm$20 & 1.07 & SL2S & 1.00 & 0.27 & 179$\pm$22 \\
J2219-0017 & 0.289 & 1.020 & 189$\pm$20 & 0.52 & SL2S & 0.70 & 1.01 & 191$\pm$20 \\
J2220+0106 & 0.232 & 1.070 & 127$\pm$15 & 2.16 & SL2S & 1.00 & 0.80 & 132$\pm$16 \\
J2221+0115 & 0.325 & 2.350 & 222$\pm$23 & 1.40 & SL2S & 1.00 & 1.12 & 227$\pm$24 \\
J2222+0012 & 0.436 & 1.360 & 221$\pm$22 & 1.44 & SL2S & 1.00 & 1.56 & 223$\pm$22 \\
J0008-0004 & 0.440 & 1.192 & 193$\pm$36 & 1.16 & SLACS & 1.50 & 1.71 & 197$\pm$37 \\
J0029-0055 & 0.227 & 0.931 & 229$\pm$18 & 0.96 & SLACS & 1.50 & 2.16 & 232$\pm$18 \\
J0037-0942 & 0.196 & 0.632 & 279$\pm$10 & 1.53 & SLACS & 1.50 & 2.19 & 283$\pm$10 \\
J0044+0113 & 0.120 & 0.196 & 266$\pm$13 & 0.79 & SLACS & 1.50 & 2.61 & 267$\pm$13 \\
J0109+1500 & 0.294 & 0.525 & 251$\pm$19 & 0.69 & SLACS & 1.50 & 1.38 & 259$\pm$20 \\
J0157-0056 & 0.513 & 0.924 & 295$\pm$47 & 0.79 & SLACS & 1.50 & 1.06 & 308$\pm$49 \\
J0216-0813 & 0.332 & 0.524 & 333$\pm$23 & 1.16 & SLACS & 1.50 & 2.67 & 335$\pm$23 \\
J0252+0039 & 0.280 & 0.982 & 164$\pm$12 & 1.04 & SLACS & 1.50 & 1.39 & 169$\pm$12 \\
J0330-0020 & 0.351 & 1.071 & 212$\pm$21 & 1.10 & SLACS & 1.50 & 1.20 & 220$\pm$22 \\
J0405-0455 & 0.075 & 0.810 & 160$\pm$8 & 0.80 & SLACS & 1.50 & 1.36 & 165$\pm$8 \\
J0728+3835 & 0.206 & 0.688 & 214$\pm$11 & 1.25 & SLACS & 1.50 & 1.78 & 219$\pm$11 \\
J0737+3216 & 0.322 & 0.581 & 338$\pm$17 & 1.00 & SLACS & 1.50 & 2.82 & 339$\pm$17 \\
J0808+4706 & 0.219 & 1.025 & 236$\pm$11 & 1.23 & SLACS & 1.50 & 2.42 & 238$\pm$11 \\
J0822+2652 & 0.241 & 0.594 & 259$\pm$15 & 1.17 & SLACS & 1.50 & 1.82 & 264$\pm$15 \\
J0841+3824 & 0.116 & 0.657 & 225$\pm$11 & 1.41 & SLACS & 1.50 & 4.21 & 222$\pm$11 \\
J0903+4116 & 0.430 & 1.065 & 223$\pm$27 & 1.29 & SLACS & 1.50 & 1.78 & 228$\pm$28 \\
J0912+0029 & 0.164 & 0.324 & 326$\pm$12 & 1.63 & SLACS & 1.50 & 3.87 & 323$\pm$12 \\
J0935-0003 & 0.348 & 0.467 & 396$\pm$35 & 0.87 & SLACS & 1.50 & 4.24 & 391$\pm$35 \\
J0936+0913 & 0.190 & 0.588 & 243$\pm$12 & 1.09 & SLACS & 1.50 & 2.11 & 246$\pm$12 \\
J0946+1006 & 0.222 & 0.608 & 263$\pm$21 & 1.38 & SLACS & 1.50 & 2.35 & 266$\pm$21 \\
J0956+5100 & 0.240 & 0.470 & 334$\pm$17 & 1.33 & SLACS & 1.50 & 2.19 & 338$\pm$17 \\
J0959+0410 & 0.126 & 0.535 & 197$\pm$13 & 0.99 & SLACS & 1.50 & 1.39 & 203$\pm$13 \\
J1016+3859 & 0.168 & 0.439 & 247$\pm$13 & 1.09 & SLACS & 1.50 & 1.46 & 254$\pm$13 \\
J1020+1122 & 0.282 & 0.553 & 282$\pm$18 & 1.20 & SLACS & 1.50 & 1.59 & 289$\pm$18 \\
J1023+4230 & 0.191 & 0.696 & 242$\pm$15 & 1.41 & SLACS & 1.50 & 1.77 & 247$\pm$15 \\
J1100+5329 & 0.317 & 0.858 & 187$\pm$23 & 1.52 & SLACS & 1.50 & 2.24 & 189$\pm$23 \\
J1106+5228 & 0.096 & 0.407 & 262$\pm$13 & 1.23 & SLACS & 1.50 & 1.68 & 268$\pm$13 \\
J1112+0826 & 0.273 & 0.630 & 320$\pm$20 & 1.49 & SLACS & 1.50 & 1.50 & 329$\pm$21 \\
J1134+6027 & 0.153 & 0.474 & 239$\pm$12 & 1.10 & SLACS & 1.50 & 2.02 & 243$\pm$12 \\
J1142+1001 & 0.222 & 0.504 & 221$\pm$22 & 0.98 & SLACS & 1.50 & 1.91 & 225$\pm$22 \\
J1143-0144 & 0.106 & 0.402 & 269$\pm$13 & 1.68 & SLACS & 1.50 & 4.80 & 264$\pm$13 \\
J1153+4612 & 0.180 & 0.875 & 226$\pm$15 & 1.05 & SLACS & 1.50 & 1.16 & 235$\pm$16 \\
J1204+0358 & 0.164 & 0.631 & 267$\pm$17 & 1.31 & SLACS & 1.50 & 1.47 & 275$\pm$17 \\
J1205+4910 & 0.215 & 0.481 & 281$\pm$14 & 1.22 & SLACS & 1.50 & 2.59 & 283$\pm$14 \\
J1213+6708 & 0.123 & 0.640 & 292$\pm$15 & 1.42 & SLACS & 1.50 & 3.23 & 291$\pm$15 \\
J1218+0830 & 0.135 & 0.717 & 219$\pm$11 & 1.45 & SLACS & 1.50 & 3.18 & 218$\pm$11 \\
J1250+0523 & 0.232 & 0.795 & 252$\pm$14 & 1.13 & SLACS & 1.50 & 1.81 & 257$\pm$14 \\
J1251-0208 & 0.224 & 0.784 & 233$\pm$23 & 0.84 & SLACS & 1.50 & 2.61 & 234$\pm$23 \\
J1330-0148 & 0.081 & 0.712 & 185$\pm$9 & 0.87 & SLACS & 1.50 & 0.89 & 194$\pm$9 \\
J1402+6321 & 0.205 & 0.481 & 267$\pm$17 & 1.35 & SLACS & 1.50 & 2.70 & 268$\pm$17 \\
J1403+0006 & 0.189 & 0.473 & 213$\pm$17 & 0.83 & SLACS & 1.50 & 1.46 & 219$\pm$17 \\
J1416+5136 & 0.299 & 0.811 & 240$\pm$25 & 1.37 & SLACS & 1.50 & 1.43 & 247$\pm$26 \\
J1430+4105 & 0.285 & 0.575 & 322$\pm$32 & 1.52 & SLACS & 1.50 & 2.55 & 324$\pm$32 \\
J1436-0000 & 0.285 & 0.805 & 224$\pm$17 & 1.12 & SLACS & 1.50 & 2.24 & 227$\pm$17 \\
J1451-0239 & 0.125 & 0.520 & 223$\pm$14 & 1.04 & SLACS & 1.50 & 2.48 & 225$\pm$14 \\
J1525+3327 & 0.358 & 0.717 & 264$\pm$26 & 1.31 & SLACS & 1.50 & 2.90 & 264$\pm$26 \\
J1531-0105 & 0.160 & 0.744 & 279$\pm$14 & 1.71 & SLACS & 1.50 & 2.50 & 281$\pm$14 \\
J1538+5817 & 0.143 & 0.531 & 189$\pm$12 & 1.00 & SLACS & 1.50 & 1.58 & 194$\pm$12 \\
J1621+3931 & 0.245 & 0.602 & 236$\pm$20 & 1.29 & SLACS & 1.50 & 2.14 & 239$\pm$20 \\
J1627-0053 & 0.208 & 0.524 & 290$\pm$14 & 1.23 & SLACS & 1.50 & 1.98 & 295$\pm$14 \\
J1630+4520 & 0.248 & 0.793 & 276$\pm$16 & 1.78 & SLACS & 1.50 & 1.96 & 281$\pm$16 \\
J1636+4707 & 0.228 & 0.674 & 231$\pm$15 & 1.09 & SLACS & 1.50 & 1.68 & 236$\pm$15 \\
J2238-0754 & 0.137 & 0.713 & 198$\pm$11 & 1.27 & SLACS & 1.50 & 2.33 & 200$\pm$11 \\
J2300+0022 & 0.228 & 0.464 & 279$\pm$17 & 1.24 & SLACS & 1.50 & 1.83 & 285$\pm$17 \\
J2303+1422 & 0.155 & 0.517 & 255$\pm$16 & 1.62 & SLACS & 1.50 & 3.28 & 254$\pm$16 \\
J2321-0939 & 0.082 & 0.532 & 249$\pm$8 & 1.60 & SLACS & 1.50 & 4.11 & 246$\pm$8 \\
J2341+0000 & 0.186 & 0.807 & 207$\pm$13 & 1.44 & SLACS & 1.50 & 3.15 & 207$\pm$13 \\
J0143-1006 & 0.221 & 1.1046 & 203$\pm$17 & 1.23 & SLACS2017 & 1.50 & 3.24 & 202$\pm$17 \\
J0159-0006 & 0.1584 & 0.7477 & 216$\pm$18 & 0.92 & SLACS2017 & 1.50 & 1.58 & 222$\pm$18 \\
J0324+0045 & 0.321 & 0.9199 & 183$\pm$19 & 0.55 & SLACS2017 & 1.50 & 1.67 & 187$\pm$19 \\
J0324-0110 & 0.4456 & 0.6239 & 310$\pm$38 & 0.63 & SLACS2017 & 1.50 & 2.23 & 314$\pm$38 \\
J0753+3416 & 0.1371 & 0.9628 & 208$\pm$12 & 1.23 & SLACS2017 & 1.50 & 1.89 & 212$\pm$12 \\
J0754+1927 & 0.1534 & 0.7401 & 193$\pm$16 & 1.04 & SLACS2017 & 1.50 & 1.46 & 199$\pm$16 \\
J0757+1956 & 0.1206 & 0.8326 & 206$\pm$11 & 1.62 & SLACS2017 & 1.50 & 3.67 & 204$\pm$11 \\
J0826+5630 & 0.1318 & 1.2907 & 163$\pm$8 & 1.01 & SLACS2017 & 1.50 & 1.64 & 167$\pm$8 \\
J0847+2348 & 0.1551 & 0.5327 & 199$\pm$16 & 0.96 & SLACS2017 & 1.50 & 1.54 & 204$\pm$16 \\
J0851+0505 & 0.1276 & 0.6371 & 175$\pm$11 & 0.91 & SLACS2017 & 1.50 & 1.35 & 181$\pm$11 \\
J0920+3028 & 0.2881 & 0.3918 & 297$\pm$17 & 0.70 & SLACS2017 & 1.50 & 4.25 & 293$\pm$17 \\
J0955+3014 & 0.3214 & 0.4671 & 271$\pm$33 & 0.54 & SLACS2017 & 1.50 & 2.95 & 271$\pm$33 \\
J0956+5539 & 0.1959 & 0.8483 & 188$\pm$11 & 1.17 & SLACS2017 & 1.50 & 1.96 & 191$\pm$11 \\
J1010+3124 & 0.1668 & 0.4245 & 221$\pm$11 & 1.14 & SLACS2017 & 1.50 & 3.26 & 220$\pm$11 \\
J1031+3026 & 0.1671 & 0.7469 & 197$\pm$13 & 0.88 & SLACS2017 & 1.50 & 1.04 & 206$\pm$14 \\
J1040+3626 & 0.1225 & 0.2846 & 186$\pm$10 & 0.59 & SLACS2017 & 1.50 & 1.30 & 192$\pm$10 \\
J1041+0112 & 0.1006 & 0.2172 & 200$\pm$7 & 0.60 & SLACS2017 & 1.50 & 2.50 & 201$\pm$7 \\
J1048+1313 & 0.133 & 0.6679 & 195$\pm$10 & 1.18 & SLACS2017 & 1.50 & 1.90 & 199$\pm$10 \\
J1051+4439 & 0.1634 & 0.538 & 216$\pm$16 & 0.99 & SLACS2017 & 1.50 & 1.66 & 221$\pm$16 \\
J1056+4141 & 0.1343 & 0.8318 & 157$\pm$10 & 0.72 & SLACS2017 & 1.50 & 1.81 & 160$\pm$10 \\
J1101+1523 & 0.178 & 0.5169 & 270$\pm$15 & 1.18 & SLACS2017 & 1.50 & 0.89 & 283$\pm$16 \\
J1116+0729 & 0.1697 & 0.686 & 190$\pm$11 & 0.82 & SLACS2017 & 1.50 & 2.44 & 192$\pm$11 \\
J1127+2312 & 0.1303 & 0.361 & 230$\pm$9 & 1.25 & SLACS2017 & 1.50 & 2.69 & 231$\pm$9 \\
J1137+1818 & 0.1241 & 0.4627 & 222$\pm$8 & 1.29 & SLACS2017 & 1.50 & 1.79 & 227$\pm$8 \\
J1142+2509 & 0.164 & 0.6595 & 159$\pm$10 & 0.79 & SLACS2017 & 1.50 & 1.51 & 163$\pm$10 \\
J1144+0436 & 0.1036 & 0.2551 & 207$\pm$14 & 0.76 & SLACS2017 & 1.50 & 1.22 & 215$\pm$15 \\
J1213+2930 & 0.0906 & 0.5954 & 232$\pm$7 & 1.35 & SLACS2017 & 1.50 & 1.73 & 237$\pm$7 \\
J1301+0834 & 0.0902 & 0.5331 & 178$\pm$8 & 1.00 & SLACS2017 & 1.50 & 1.25 & 184$\pm$8 \\
J1330+1750 & 0.2074 & 0.3717 & 250$\pm$12 & 1.01 & SLACS2017 & 1.50 & 2.85 & 251$\pm$12 \\
J1403+3309 & 0.0625 & 0.772 & 190$\pm$6 & 1.02 & SLACS2017 & 1.50 & 2.00 & 193$\pm$6 \\
J1430+6104 & 0.1688 & 0.6537 & 180$\pm$15 & 1.00 & SLACS2017 & 1.50 & 2.24 & 182$\pm$15 \\
J1433+2835 & 0.0912 & 0.4115 & 230$\pm$6 & 1.53 & SLACS2017 & 1.50 & 3.23 & 229$\pm$6 \\
J1541+3642 & 0.1406 & 0.7389 & 194$\pm$11 & 1.17 & SLACS2017 & 1.50 & 1.55 & 199$\pm$11 \\
J1543+2202 & 0.2681 & 0.3966 & 285$\pm$16 & 0.78 & SLACS2017 & 1.50 & 2.32 & 288$\pm$16 \\
J1550+2020 & 0.1351 & 0.3501 & 243$\pm$9 & 1.01 & SLACS2017 & 1.50 & 1.68 & 249$\pm$9 \\
J1553+3004 & 0.1604 & 0.5663 & 194$\pm$15 & 0.84 & SLACS2017 & 1.50 & 2.15 & 197$\pm$15 \\
J1607+2147 & 0.2089 & 0.4865 & 197$\pm$16 & 0.57 & SLACS2017 & 1.50 & 2.63 & 198$\pm$16 \\
J1633+1441 & 0.1281 & 0.5804 & 231$\pm$9 & 1.39 & SLACS2017 & 1.50 & 2.39 & 233$\pm$9 \\
J2309-0039 & 0.2905 & 1.0048 & 184$\pm$13 & 1.14 & SLACS2017 & 1.50 & 2.08 & 187$\pm$13 \\
J2324+0105 & 0.1899 & 0.2775 & 245$\pm$15 & 0.59 & SLACS2017 & 1.50 & 1.10 & 255$\pm$16 \\
&& \\
\hline\hline
\end{longtable}
\end{center}

Let us now describe the sequence of steps taken to minimize the overall scatter
in the data. If we simply use the full set of 158 confirmed sources, without the 
introduction of an additional dispersion associated with the SIS and the exclusion
of outliers, we find that {\it Planck} $\Lambda$CDM (with $\Omega_{\rm m}=0.308$, 
$\Omega_{\rm de}\equiv \Omega_\Lambda=1-\Omega_{\rm m}$) fits the strong lenses 
with a reduced $\chi^2_{\rm dof}$ ($\chi^2$ per degree of freedom) of 
$\backsimeq2.7$, which is not satisfactory. A more serious issue 
is that most of the data---89 out of 158 sources---are inconsistent with Planck 
$\Lambda$CDM at more than $1\sigma$. By comparison, if the reported errors were truly 
Gaussian, we should expect $\approx 50$ of the 158 measurements to deviate by
more than $1\sigma$ from an accurate cosmological model. But there are clearly
several mitigating circumstances. For example, in our initial analysis, we find that
a single source, J0850-0347, deviates by more than $5\sigma$ from all the considered 
models, and we therefore exclude it as a significant outlier from all further 
consideration. With this single source excluded, the  $\chi^2_{\rm dof}$ for
{\it Planck} $\Lambda$CDM immediately drops to $\backsimeq2.4$.

This improvement notwithstanding, such a poor $\chi^2_{\rm dof}$ contrasts 
sharply with the outcome reported in Melia, Wei \& Wu (2015), but we note that an 
additional error term, $\sigma_{\rm f}$, was included in that earlier paper to 
characterize possible random deviations from the simple isothermal sphere 
(SIS) model. This dispersion was assumed to be $6\%$, resulting in a $12\%$ 
contribution to $\sigma_\mathcal{D}$ based on standard error propagation. 
Nonetheless, were we to include that additional scatter here, the resulting 
$\chi^2_{\rm dof}{\backsimeq}1.5$ for the {\it Planck} $\Lambda$CDM best fit 
would still be significantly greater than the value (i.e., $\chi^2_{\rm dof}{\backsimeq}1.2$) 
found in our previous work. The difference is entirely due to the new data we
have added to the sample in this paper, at least some of which appear to deviate 
significantly from the {\it Planck} model.

We believe that contributing factors to this disparity are (i) that the reported 
errors are possibly underestimated, and (ii) that there is an additional unrecognized 
systematic effect that has yet to be included in the analysis. In addition, we carefully
excluded from our previous analysis those lens systems with $\mathcal{D^{\rm obs}}>1$, 
which are unphysical (see Eq.~8). If we follow the same steps here with the larger 
sample, introducing the additional SIS dispersion and excluding the 28 lenses with 
$\mathcal{D^{\rm obs}}>1$, 11 of which are from the 2017 SLACS catalog, and also 
exclude the aforementioned extreme outlier J0850-0347, we find that
$\chi^2_{\rm dof}\approx 1.01$ for the {\it Planck} $\Lambda$CDM cosmology, nearly 
identical to our previous result. As we explain in more detail below, there are
good reasons for believing that the SIS dispersion may be bigger than the value
we used previously. For example, in their fitting, Cao et al. (2012) invoked possible 
deviations from SIS contributing a scatter of up to $20\%$. This appears to be
more in line with our preliminary finding here, so we investigate the impact of 
such a large dispersion on our optimization of the parameters in the $w$CDM cosmology.

We use maximization of the likelihood function to constrain and compare the models,
including $w$CDM. We calculate $\mathcal{D}^{\rm obs}$ using Equation~(3), and 
$\mathcal{D}^{\rm th}$ using Equation~(7). For each measurement of 
$\mathcal{D}^{\rm obs}$ we also determine the corresponding error
through standard error propagation, in which
\begin{equation}
\sigma_{\mathcal{D}^{\rm obs}}=\mathcal{D}^{\rm obs}\,
\sqrt{\left(\dfrac{\sigma_{\theta_E}}{\theta_E}\right)^2+
\left(\dfrac{2\sigma_{\sigma_0}}{\sigma_0}\right)^2+\sigma_X^2}\;,
\end{equation}
where $\sigma_X$ is a unitless composite error term comprising the scatter 
about the SIS average and any other source of scatter in the measurements. 
We iterate the value of $\sigma_X$ (described below), while also systematically 
eliminating sources with a $\mathcal{D}^{\rm obs}$ exceeding $1$,
since these are clearly unphysical, and we also exclude J0850-0347, 
which is an extreme outlier in every model we tested.

In the method of maximum likelihood estimation (MLE; Wei at al. 2015a),
the joint likelihood function for all parameters, based on a flat Bayesian 
prior, is
\begin{equation}
\mathcal{L}=\prod_i\dfrac{1}{\sqrt{2\pi}\,\sigma_{\mathcal{D}_i}}\;
\exp\left[{-\dfrac{\chi^2_i}{2}}\right]
\end{equation}
where, for each measurement,
\begin{equation}
\chi_i^2\equiv\dfrac{\left(\mathcal{D}_i^{\rm obs}-
\mathcal{D}_i^{\rm th}\right)^2}{\sigma_{\mathcal{D},i}^2}\;.
\end{equation}
The iteration on $\sigma_X$ ends when the optimization of $w$CDM results
in a $\chi^2_{\rm dof}=1$. Once $\sigma_X$ has been identified in this
way, we use the same value for all the models in order to keep the 
comparison as transparent as possible. Although this approach tends to
favour $w$CDM somewhat, we will see that it does not 
influence the model ranking significantly. For example, the disfavoured
models are rejected strongly and, clearly, changing $\sigma_X$ by a few
percentage points will not alter this outcome.

\vskip 0.1in
\begin{table}
  \large
  \caption{Best-fitting Parameters for the CDM models (Constrained by $\mathcal{D}^{\rm obs}\le 1$)}
  \centering
  \begin{tabular}{lcccc}
&& \\
    \hline
\hline
&& \\
Model & $\Omega_{\rm m}$ & $w_{\rm de}$ & $\sigma_X$ & $\chi^2_{\rm dof}$\\
\hline
&&\\
$wCDM$&$0.33_{-0.15}^{+0.13}$ & $-1.29_{-6.09}^{+0.97}$&12.2\%&0.998 \\
$\Lambda$$CDM$&$0.29_{-0.08}^{+0.12}$&$-1$ (Fixed)&12.2\%&0.999 \\
&& \\
\hline\hline
  \end{tabular}
\end{table}
\vskip 0.1in

\begin{figure}
\begin{center}
\includegraphics[width=0.5\linewidth]{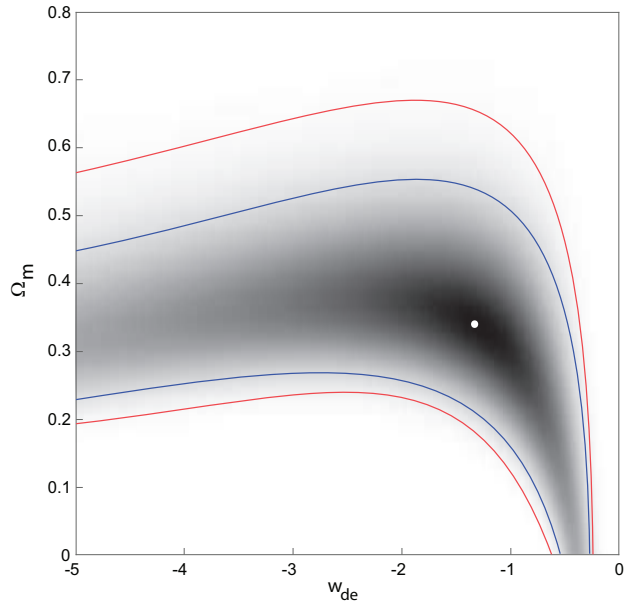}
\end{center}
\caption{Probability density plot in the $\Omega_{\rm m}-w_{\rm de}$ plane
for $w$CDM. The contours give the $1\sigma$ and $2\sigma$ (i.e., the $68\%$ and 
$95\%$) confidence regions for the optimized parameters in $w$CDM. The white dot
shows the best fit values.}
\end{figure}

The resulting best-fit values (including $\sigma_X$), calculated via the 
marginal probability of each parameter, are reported in Table~2 for both
$w$CDM and $\Lambda$CDM. The latter has one fewer parameter (since 
$w_{\rm de}=-1$). The corresponding $1\sigma$ and $2\sigma$ confidence 
contours for $w$CDM in the $\Omega_{\rm m}-w_{\rm de}$ plane, along with 
the overall probability density, are shown in fig.~1. Retaining only
sources with $\mathcal{D^{\rm obs}}\le 1$ reduces our overall sample size 
from 157 (excluding the extreme outlier) to 129 lenses, but it also 
decreases the magnitude of $\sigma_X$ compared to
what it would have been for the entire sample. A quick inspection of
Table~2 shows that the steps we have taken in identifying the final
sample for model selection produces results that are consistent with
the {\it Planck} measurements. Both $\Omega_{\rm m}$ and $w_{\rm de}$
are fully consistent with the parameter values in the concordance
model, particularly in the case of $\Lambda$CDM.

With the best-fitting parameters for flat $w$CDM thus determined, we now
proceed to carry out model selection based on the Bayes information criterion 
(BIC; Schwarz 1978; Melia \& Maier 2013; Wei et al. 2015b). The BIC is defined as 
\begin{equation}
\text{BIC}=-2\mathcal{L}+n\ln(N)\;, 
\end{equation}
where $\mathcal{L}$ is the likelihood in Equation~(11), $N$ is the number 
of measurements in the final reduced sample, here 129, and $n$ is the number 
of free parameters. In this application, $w$CDM is penalized with $n=2$, and
$\Lambda$CDM with $n=1$, while $R_{\rm h}=ct$, Milne, and Einstein-de-Sitter 
each have $n=0$ (no free parameters).  When comparing cosmologies using the 
BIC, the  probability that a specific model $\mathcal{M}_\alpha$ is the correct 
one among the set being considered is
\begin{equation}
P(\mathcal{M}_\alpha)=\dfrac{\exp(-\text{BIC}_\alpha/2)}{\sum_i\exp(-\text{BIC}_i/2)}\;.
\end{equation}
Table~3 summarizes the $\chi^2_{\rm dof}$, the BIC, and {\it relative} likelihood 
(calculated from Eq.~14) of each model in this comparison.

\vskip 0.1in
\begin{table}
 \small
  \caption{Model Comparisons and Ranking (based on 129 lenses with the 
constraint $\mathcal{D}^{obs}\le1$)} 
  \centering
  \begin{tabular}{lcccc}
&& \\
    \hline
\hline
&& \\
Model & $\chi^2_{\rm dof}$&BIC&Relative Likelihood\\
\hline
&&\\
$R_{\rm h}=ct$ & $1.020$ &$131.559$&73.032\% \\
$\Lambda$CDM & 0.999 & 133.748 & 24.443\% \\
$w$CDM & $0.998$ &$138.484$&2.290\% \\
Milne& $1.109$ &$143.063$&$0.232\%$ \\
EdS & $1.194$ &$151.314$&$3.75{\times}10^{-3}\% $\\
&& \\
\hline\hline
  \end{tabular}
\end{table}
\vskip 0.1in

\section{Discussion}
Our analysis in this paper affirms the important role played 
by strong lenses in helping to refine the parameters in the
standard model and, perhaps more importantly going forward, 
providing ample confirmation, if not definitive evidence, in 
model selection. Previous work by Melia, Wei \& Wu (2015) and 
Cao et al. (2015), albeit with smaller samples, 
suggested that---while individual lenses may deviate from 
an SIS model---the statistics of a large sample appears 
to be consistent with this simple internal structure of the lens's
mass distribution. We have therefore adopted this approach
to update the optimization of  parameters in the standard
model based on fits to the strong lens angular diameter
distance dependence on redshift, and then to compare the
predictions of $w$CDM with those of four other cosmologies.
We have found, however, that ignoring individual variations
from a pure SIS structure results in an unsatisfactory fit
using $w$CDM and $\Lambda$CDM, necessitating the introduction
of a dispersion to represent the scatter associated with
this over-simplified lens model.

It is important to emphasize that our sample is larger
than that used in any previous attempt to carry out this
type of analysis, and that it includes all of the sources
used by Cao et al. (2015) (with the exception of a single outlier).
We have supplemented this catalog with the 40 recently confirmed
lenses uncovered with SLACS (Shu et al. 2017). Our best-fit
parameters for the standard model are consistent with those
of {\it Planck}, but based solely on the reported errors, the
reduced $\chi^2_{\rm dof}$ for the optimized model is unacceptably
large, unless we include the aforementioned additional scatter in 
the analysis. We argue that either the errors have been under-reported, 
or that the additional dispersion cannot be ignored with observations 
such as these. We have therefore sought to identify its magnitude,
representing deviations from a pure homogeneous SIS model for all 
the lenses. Note, however, that 
our inferred uncertainty on the optimized value of $w_{\rm de}$ 
(in the case of $w$CDM) is larger than that obtained by Cao 
et al. (2015), in spite of the larger lens catalog at our disposal.

\begin{figure}
\begin{center}
\includegraphics[width=0.8\linewidth]{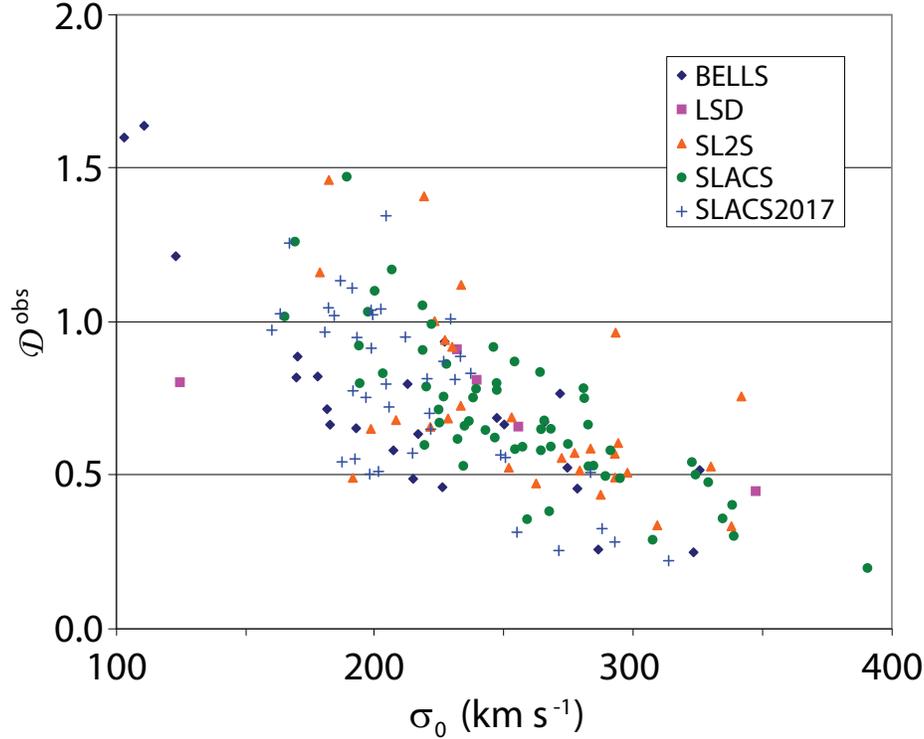}
\end{center}
\caption{An apparent bias in the measured ratio $\mathcal{D}^{obs}$ 
with increasing velocity dispersion $\sigma_0$. This trend exceeds the
dependence of $\mathcal{D}^{obs}$ on $\sigma_0$ expected from Eq.~(3),
in which $\theta_{\rm E}$ ought to change in concert with $\sigma_0$
to largely offset such a correlation. Two outliers, J2220+0106 and 
J1352+3216, lie outside this plot due to their extreme $\mathcal{D}^{obs}$ 
values of 4.32 and 2.36, respectively.}
\end{figure}

But though the introduction of the dispersion $\sigma_X$ to account
for individual departures from a pure SIS lens structure has greatly
reduced the scatter about the best-fit model in $w$CDM/$\Lambda$CDM, 
the problems with using such a simple lens have not been completely 
eliminated, as demonstrated in fig.~2. Cao et al. (2015) reported a 
significant trend in deviations from their fitted cosmological model 
as a function of $\sigma_0$. We find the same trend with the larger
sample used in this paper, in which $\mathcal{D}^{obs}$ decreases faster
than expected from Eq.~(3) with increasing $\sigma_0$. Certainly 
Eq.~(3) predicts that for a fixed $\theta_{\rm E}$, one should see
such a trend. But the Einstein radius also depends on the lens 
galaxy's mass distribution and the distance ratio between the lens 
and source objects, which should largely offset the trend seen
in fig.~2. There is no reason to expect a significant correlation 
between the mass distribution of the lensing object and the distance 
ratio between it and the source galaxy, as these are two 
independent parameters that produce the Einstein ring. For this reason, 
such a significant correlation between $\mathcal{D^{\rm obs}}$ and
$\sigma_0$ can be taken as some evidence that the simple SIS
galaxy mass distribution model is not robust enough to accurately 
account for all the individual variations seen from source to source. 

In the redshift range of these data, the best-fit $w$CDM, 
$\Lambda$CDM and $R_{\rm h}=ct$ models predict comparable 
$\mathcal{D^{\rm th}}$ ratios. In each case, strongly lensed 
systems with  $\sigma_0 \lesssim 250$ km s$^{-1}$ generally have 
$\mathcal{D}^{\rm obs}>\mathcal{D}^{\rm th}$, but this trend
is reversed for $\sigma_0 \gtrsim 250$ km s$^{-1}$. The effect
tends to get bigger as $\sigma_0$ increases or decreases away
from its median value $\sim 233$ km s$^{-1}$. In fact, all 
lens systems with an unphysical $\mathcal{D}^{\rm obs}>1$ have 
$\sigma_0 \lesssim 233$ km s$^{-1}$. Working with a smaller sample
of the data than we have here, Cao et al. (2015) attempted to
generalize the SIS model by characterizing it as a spherically 
symmetric power-law mass distribution of the form 
$\rho{\sim}r^{-\gamma}$. Their optimized value of $\gamma$ was 
consistent with $-2$, however, which is in fact the SIS model, 
but they also noticed large deviations from their fits for 
$\sigma_0$ very different from 250 km s$^{-1}$. One should
therefore be cautious with the use of an SIS model in 
future attempts to constrain or compare cosmologies using
strong lensing data. At a minimum, one should carefully study
the impact of a density profile varying with changing $\sigma_0$.

A principal goal of this paper has been to significantly update the 
results of Melia, Wei \& Wu (2015). In that analysis, with only 69 
strong lenses, no significant preference was determined for either
$\Lambda$CDM or $R_{\rm h}=ct$. Based on a much larger sample of mock data,
however, these authors concluded that approximately 200 lenses would 
be required to show that $\Lambda$CDM is preferred over $R_{\rm h}=ct$ 
at the $99.7\%$ confidence level if the standard model is the correct
description of nature. On the flip side, this earlier work also showed
that about 300 lenses would be needed to demonstrate the superiority of
$R_{\rm h}=ct$ over $\Lambda$CDM at a comparable level of confidence if
it turned out that the former was the correct model. With the 158 systems
we have considered here, reduced to 129 with the exclusion of the unphysical
ones, our sample has not quite reached that size yet, but we are rapidly
approaching these thresholds. As Table~3 shows, the current
status has the $R_{\rm h}=ct$ universe as the preferred model, followed
by $\Lambda$CDM, which is favoured over $w$CDM. Although $w$CDM is 
slightly more flexible in fitting the data than $\Lambda$CDM, the
penalty incurred by the additional free parameter causes it to
be favoured less than the standard model with a fixed dark-energy
equation of state. At the same time, these results show that the
Milne universe and Einstein-de Sitter are completely ruled out. The
sample of strong lenses now available for model selection is therefore
already large enough to provide results consistent with those of many
other kinds of observation, all of which have thus far tended to favour
$R_{\rm h}=ct$ over $w$CDM/$\Lambda$CDM (see, e.g., Melia 2013a, 2013b; 
and especially Table~1 in Melia 2017b).

\section{Conclusion} 
An important byproduct of this analysis has been our assessment
of the likely intrinsic scatter associated with the SIS model for the 
lens. If the random variation in galaxy morphology is almost Gaussian, 
we find that an additional error term of about $12.22\%$ is necessary 
to have 68\% of the observations lie within $1\sigma$ of the best-fit 
$w$CDM model. This factor is smaller than---though consistent 
with---the $20\%$ scatter suggested by Cao et al. (2012). Thus,
in spite of the fact that our sample here is twice as large as
that used in our previous analysis (Melia, Wei \& Wu 2015), 
our conclusion regarding the size of this scatter is virtually 
identical to that of our previous work, in which we found that
$\sigma_X\sim 0.12$. Interestingly, this is is very close to
the conclusion drawn earlier by Treu et al. (2006), 
who also argued for the inclusion of a scatter of about $12\%$. 
But this is only true when sources with $\mathcal{D}^{\rm obs}>1$
are excluded. Were we to include all 157 sources (the complete
sample of 158 minus the significant outlier J0850-0347), we
would find that $\sigma_X$ is closer to $18\%$. As noted, this
difference provides some evidence that the SIS lens model breaks
down for the more extreme values of $\sigma_0$.

Based on our earlier work (Melia, Wei \& Wu 2015) and the significant
improvement we have seen using a much bigger sample in this paper, we 
are certain that strong lenses will play a pivotal role in model
selection going forward---but preferably with an improved model
for the lens mass. Already DES has released a catalog of 348 
new strong lens candidates (Diehl et al. 2017). Spectroscopic follow-up 
observations are anticipated over the next several years. Even if 
only half of these are verified lenses, with a sufficient number of 
sources at $z>3$, we anticipate that the next update of our analysis 
may offer an even stronger answer as to whether $R_{\rm h}=ct$ or 
$\Lambda$CDM is the correct cosmology.

\section*{Acknowledgments}
We are grateful to Joel Brownstein, Raphel Gavazzi, and Tommaso Treu for their 
correspondence and helpful suggestions, and to the anonymous referee for
a very helpful and thoughtful review. FM is supported by the Chinese Academy of 
Sciences Visiting Professorships for Senior International Scientists under grant 
2012T1J0011, and the Chinese State Administration of Foreign Experts Affairs under 
grant GDJ20120491013.

\label{lastpage}

\end{document}